# Interacting with Masculinities:
# A Scoping Review

**KATIE SEABORN**

*Mechanical & Industrial Engineering*
*University of Toronto*

**ABSTRACT:** Gender is a hot topic in the field of human-computer interaction (HCI). Work has run the gamut, from assessing how we embed gender in our computational creations to correcting systemic sexism, online and off. While gender is often framed around women and femininities, we must recognize the genderful nature of humanity, acknowledge the evasiveness of men and masculinities, and avoid burdening women and genderful folk as the central actors and targets of change. Indeed, critical voices have called for a shift in focus to masculinities, not only in terms of privilege, power, and patriarchal harms, but also participation, plurality, and transformation. To this end, I present a 30-year history of masculinities in HCI work through a scoping review of 126 papers published to the ACM Human Factors in Computing Systems (CHI) conference proceedings. I offer a primer and agenda grounded in the CHI and extant literatures to direct future work.

**KEYWORDS:** Masculinities, masculinity, gender, CHI, human-computer interaction, scoping review, feminist HCI

# 1  Introduction

Gender is on the frontlines of work aiming to raise attention to matters of inclusion, diversity, and social justice in human-computer interaction (HCI) [23,135,145,178,180,191]. Gender is a multifaceted aspect of human identity and social organization [60,69,95,96]. In line with emerging academic and cross-cultural consensus [95,96], I approach gender as a social construct that is constituted, negotiated, and performed in a multitude of ways for a variety of functions within societies[1]. Gender can be a mode of expression, an internal self-identity, an external social category, and/or an abstract perception, even of objects [29,69,120,222]. Gender is often framed as femininity and masculinity, but a wealth of work across time and cultures has challenged this "binary" model [60,95,96,145,178,188,198], including in HCI [178,188,191]. Indeed, HCI has a history of developing, studying, and critiquing technologies for gender inclusion and anti-sexism [12,15,165]. Feminist and intersectional HCI [178] projects have appeared alongside social movements like #MeToo and organizational changes, notably the Critical Computing, Sustainability, and Social Justice subcommittee[2] at CHI [38].

Technologies, especially computer-based ones, are implicated as sites and mediums of gender. Work has explored stereotypes in design decisions [43,149,155,219], harassment in digital spaces [59,125,168], expert biases in recruitment and methods [114,121,145,180,187], and more. This work centres diversity and representation; highlights toxic behaviours towards women and genderful folk; and foregrounds the malignment of femininities and genderfulness. It has also unveiled the role of patriarchal systems [215] that centre, value, and privilege masculinities, where men and others uphold and benefit from the power offered to men by these systems. Much work has focused on limits and harms—but technology can also be a means of raising awareness and exploring gender expressions and experiences, as well.

Critical work has raised another challenge: framing [54,107,159,209,216]. When it comes to harms meted out through technology, women and genderful folk are rightfully centred. However, this can imply that sexism is under their purview alone. As Himmelsbach et al. [89:11] warn, "if women are studied solely, this might convey the impression [that gender] does not matter to men." Men and masculinities are concealed [107], with men escaping responsibility or legitimately believing that they have no role to play [77,209]. Others argue that women and genderful folk must change. A widely criticized [63,92] instantiation of this is former Meta COO Sheryl Sandberg's "lean in" feminism [174]. Some view the systems as too difficult to change or find that men do not participate [35,174], perhaps viewing sexism as a personal problem [54,209] or not knowing how to take action even when they want to [54,159]. A more subtle frame is what other masculinities can be embraced, if not toxic ones—and how technology can help.

---

[1] Gender is often contrasted with sex, which refers to the biological properties of people's bodies, such as chromosomes, hormones, sex organs, and secondary sex characteristics, e.g., facial hair, that are categorized as male, female, and/or a range of intersexes [60,95,96]. As for gender, these categories are also social constructs. Also, the properties attributed to certain categories can vary widely within and across those categories; for example, breast size.

[2] Notably, the name of the subcommittee was updated in 2022 to explicitly include the "social justice" part.

Work on masculinities in and outside of HCI may help reframe the situation and chart a path forward. A plurality of masculinities have been mapped out [107]. Masculinities intersect with other gender and sex identities [1,40], as well as other factors, such as race [26,139,189] and sexuality [106]. As a social construct [69], masculinities are in flux, co-created, contested, and concretized—wittingly or otherwise—across cultures and over time [107]. Importantly, masculinities are not set in stone, nor are they the purview of men; we must all engage in interrogative, reflective, and practical work on masculinities, as a feature of research, at least. Indeed, deconstructing and diversifying the very notion of masculinities itself will be instrumental for progress on gender equality for everyone [107,159]. We may approach masculinities as a facet of the user experience (UX), a demographic variable with a legacy of privilege within technology spaces [116,212], and/or as a designable object [19,224], one that may be shaped by computer technologies, such as social media [64,73,184]. Yet, how masculinities have been approached as a subject of study in HCI remains obscure. Moreover, most of us are not well-versed in theories of gender, as a matter of course in most forms of STEM education. This leaves us with a gap in our understanding of what has been done, what can be done, and what next steps should be taken, especially in HCI.

In this preliminary work, I sought to better understand whether and how masculinities have been approached within HCI. I asked a broad and exploratory question: *How have masculinities been approached in the field of HCI, if at all?* To this end, I carried out a scoping review [141,156,204] of the ACM Human Factors in Computing Systems (CHI) conference proceedings. I chose CHI as a comprehensive general venue featuring the highest quality of work in HCI. This work sets the stage for future primary research and systematic review work on masculinities within the field of HCI. As a retrospective, it is a stimulus for community self-reflection. It also addressing the need for more work centred on masculinities, in the plural, and how men can act in service of gender equality within HCI and the greater world.

## 2    Methods

I conducted a scoping review [141,156], a form of exploratory yet systematic literature review that aims to broadly capture a research subject, topic, or field of study [141,156]. Scoping reviews are typically carried out before systematic reviews so as to identify the extent of the available primary research so far [141,204], make a judgment on the value of carrying out systematic work [156], which has certain requirements and is much more resource-intensive [87,151], and/or summarize the findings, trajectories, and gaps, especially when the work is novel or complex [204]. The value lies in tracing out histories, clarifying concepts, identifying knowns and unknowns (or even the unasked), and mapping out next steps [141,156]. I used the PRISMA-ScR approach[3] [204], a world-class standard [151] that helps maintain rigour when carrying out review work and provides a formal structure for reporting, ideal for ease of reading and peer review, as well as later meta-review work. While I undertook this project alone, I aimed to avoid bias in my procedure by employing the PRISMA-ScR. I registered this protocol before data collection on December 31[st], 2022 at OSF[4].

### 2.1    ELIGIBILITY CRITERIA

All items, i.e., papers published to the CHI proceedings, that included masculinity as part of the work were accepted. If masculinity was referenced but not integrated into the work, e.g., in related work or future work, the item was excluded.

### 2.2    INFORMATION SOURCES AND SEARCH

The ACM Digital Library (ACM DL), the venue for the CHI proceedings, was searched on December 31[st], 2022. The search query was: *AllFields:(masculinity) OR AllField:(masculinities)*. The results were filtered to the CHI proceedings.

### 2.3    DATA ITEMS

Metadata were extracted and: topic of study; research questions (RQs); HCI context, e.g., virtual reality (VR), hackerspaces; whether masculinities were central; whether gender was central; definition(s) of masculinities; whether these were explicit and quotable, implicit, such through associations of descriptors and masculinities, or unstated; whether masculinities were approached as a plural construct; whether gender was approached as a social construct; whether a binary approach to gender was taken; types of masculinities; theories; and citation(s) for all of these.

---

[3] Note that I deviated from the PRISMA-ScR to accommodate CHI reporting structures and HCI approaches to reporting, e.g., no structured abstracts.
[4] https://osf.io/3kv7s

## 2.4    SELECTION PROCESS AND DATA SYNTHESIS

I downloaded the results of search from the ACM DL into Zotero and removed two invalid items (introductions to conference proceedings). I then randomly ordered and screened the items based on the eligibility criteria, removing five non-CHI papers. Next, I conducted a full-text review. This was done in three stages in parallel with data analysis, for which I used a reflexive thematic analysis approach [22]. This method is suitable for solo work; as Braun and Clark acknowledge, exploratory data analysis is subjective and relies on rater expertise; in my case, I am an experienced mixed methodologist in HCI who has published on gender. My process: First, I reviewed 20% of the items, and extracted data if eligible. At this stage, I developed the first set of codes based on patterns and highlights in the data. I then revisited the first 20% of items to refine the codes. I then reviewed and coded the next 20%. At this stage, I developed higher-level themes based on the expected contributions for scoping reviews but contextualized for the topic and field of HCI [141]: *social, design, research,* and *critique*. I then recoded all items. I removed 13 items that only referenced work or pointed to future work. I also categorized types of masculinities and theories. I used exploratory statistics where possible.

## 3    Results

From an initial 146 records, 126 items between 1993-2022 were included (Figure 1). The data is available on OSF[5]. I now summarize the results. Counts and percentages were calculated against the total number of papers, unless specified.

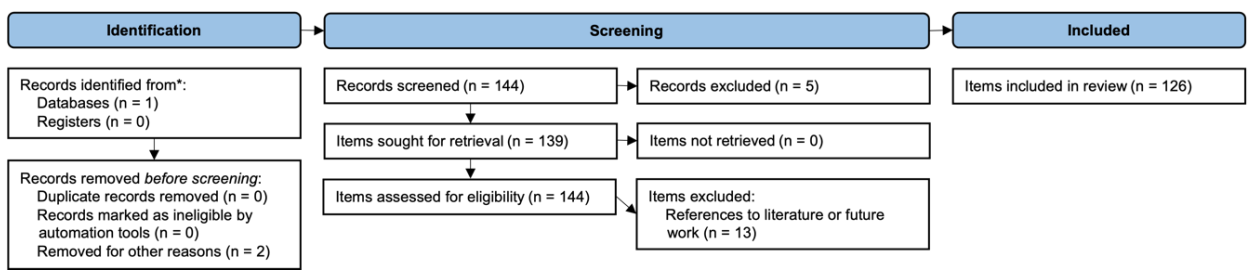

**Figure 1:** *PRISMA flow diagram[6] [151,204] of the procedure for identifying items to be included and excluded.*

---

[5] https://osf.io/qgk8m
[6] I have modified the diagram provided by Page et al. [151] to remove the medical framing and transform the figure into a horizontal format.

## 3.1 MASCULINITY-CENTRISM IN TOPICS AND CONTEXTS OF STUDY

A variety of topics and 25 contexts were found. The most common contexts were general UI research, e.g., web and physical computing (17, 13%); HCI as a field, e.g., critiques of HCI practice, experiences of working in HCI, and bias in design, research, and practice (16, 13%); social media, e.g., Twitter and Reddit (14, 11%); and communities, e.g., online peer support, hackerspaces, and specific groups (13, 10%). Others included games, field work, crowdsourcing, VR, CUIs and voice UX, HRI and HAI, and design practice. Topics were so varied that was is difficult to summarize; I recommend reviewing the list in the OSF data set. I will draw out details and key into this diversity in the following sections.

Despite the target of this work and search terms used, only 14 (11%) papers had masculinities as a main topic. Additionally, about half or 65 (52%) papers were focused on gender. Examples of topics include: gender norms and attitudes in father blogs [124]; African American men and computing identity [98]; heteronormativity and hegemonic masculinity in VR porn [227]; sexist beliefs via sexualization/objectification of avatars within gaming contexts [21]; and urinal games [133]. Topics cover a variety of masculinities, intersectionality, identity and body, and sexism.

Most other papers had masculinities as subfactors or emergent factors. Martens [131] used a scale to evaluate a novel statistics interface that included the subfactors "feminine," "masculine," and "unisex." Williams and Gilbert [223:3] contextualized their case studies against the scholarship of feminist and critical disability studies to point out the role that certain masculinities within certain cultural contexts play in power imbalances within peer review processes: "This self-invisibility is the specifically modern, European, masculine, scientific form of the virtue of modesty ... modern fantasies of objectivity have historically rendered the white male as the invisible default witness to scientific fact."

Masculinities were also approached at a meta level, in relation to their design and/or research praxis. Schechter, Egelman, and Reeder [176] explained that they used masculine and feminine pronouns in the accounts of their social-authentication system "for clarity," a choice that assumes a reader would be confused by the use of the same pronouns for different characters, as well as one that belies gender binary thinking. Tachtler et al. [196] provided a nuanced discussion about recruitment, noting that they had aimed but were unable to recruit a diverse sample, and considering valid reasons why, i.e., most unaccompanied migrant youth in and around their site were young men. Kao et al.

[103] argued that their selection of stereotypical "male" and "female" voiced avatars was meant to achieve ecological validity against the most common setups in real games, noting that "a binary view of gender is problematic" [103:7].

## 3.2 DEFINITIONS OF MASCULINITIES

Only four papers (3%) explicitly operationalized masculinity or masculinities. These were sourced from critical studies, dictionaries, cultural studies, or were unsourced. I outline them below:

- Rubin, Blackwell, and Conley in 2020 [168:2], citing Coston and Kimmel [41]: "the behaviors and expectations culturally associated with boys and men."
- Pater et al. in 2019 [153:2-3] combine "masculinity" and "male," and "male" and "man," operationalizing these as identity expression through social media, notably by way of "external appearance." They use the Oxford English Dictionary to define "male," but refer to the dictionary term of "man": "having qualities or appearance traditionally associated with men, especially strength and aggressiveness ... gender as "spectrum" [233].
- Danielescu and Christian in 2018 [44:6], citing Hofstede's "cultural dimensions," or construction of gender as a culture-wide, general attribute [90]: "Masculinity, contrasted with femininity, is sometimes expressed as 'tough vs. tender' - it quantifies how competitive a society is, and social rewards for achievement vs. cooperation."
- Dosono and Semaan in 2018 [51:5] created a thematic code for "masculinity" in the context of American Asians and Pacific Islanders (AAPIs) on Reddit, defining it as: "Critiques of qualities traditionally associated with men."

In 14 cases (11%), definitional qualities were implied by association. A word frequency analysis of 468 terms from the sentences in which "masculinity/ies" was found highlights several commonalities: *gender* (11), *feminine,* (7), *associated, home, male, men, physical, qualities, traditionally, work* (4 each), *binary, body, competitive, data, design, female, gaming, individuals, psych, strength, traits, violence* (3 each), *activities, agentic, aggressive, alternative, assertive, categories, characterize, clothing, consistent, different, domestic, dominant, emotions, express, form, guru, identifying, identity, implicit, labor, man, men's, minorities, models, people, power, pressure, sir, sport, technical, technology, traditional, trans, transmasculine, transmen, type, user, women working* (2 each). This indicates that masculinities were positioned against femininities and women. Qualities align with common views, such as strength, violence, dominance, tradition ... and technology. Still, a portion point to alternative models, intersectional factors, and trans identities.

81 papers (64%) did not define or operationalize masculinity directly or indirectly. A word frequency analysis of 825 terms from the sentences in which "masculinity/ies" was found reveals: gender (26), feminine (19), men (18), women (14), male (11), fem, participants, technology (8 each), binary, female, two (7 each), choices, Hofstede (6 each), abuse, norms, people, sex, white (5 each), avatar, boys, communities, control, create, cultural, experiences, gendered, modern, perceived, play/games, traditional (4 each), culture, dimensions, dominant, environment, hegemonic, identity, majority, males, neutral, options, others, participant, pronouns, role, social, trans, work, young (3 each). While this largely matches the above, the work of Hofstede and cultural-level operationalizations of masculinity/ies are highlighted. Moreover, there is mention of male-as-neutral and specific reference to games/play. Finally, the intersectional factors and trans identities are less present, which we might expect since these tend to be marginalized and thus at risk of being overlooked.

Nearly half or 60 papers (48%) took a pluralistic approach to masculinities. Still, 8 papers (6%) took a singular approach, and for more than half (58, 46%) it was unclear. For example, Madden et al. [126:1] cited literature on how gamers and gaming cultures have been characterized, without making it clear how they felt about it: "male and female gamers play competitive games in roughly equal numbers ... esports ... are still viewed as 'male-dominated' ... the masculine and feminine cultures in gaming are still surrounded by commonplace assumptions, such as males being 'aggressive' and 'undesirable' individuals." Merely citing literature does not necessarily imply the authors' own stances. Most papers (92, 73%) approached masculinities as a constructed object, although it was hard to judge in 32 (25%) of cases. For example, Gonzales and Fritz [76] talked to folks engaging in crowdsourcing to fund top survey or "reconstruction of a masculine chest" [76:2371], recruiting "only transmen or those who identified as transmasculine" [76:2372]. This points to plurality as well as the intersections of identity, positionality, and personal choice. Still, nearly half (58, 46%) relied on a gender binary perspective, with 54 (43%) going beyond the binary and 14 (11%) unclear. Given the clear distribution here, I ran a Kendall tau-b test, finding a significant negative relationship between year of publication and binary positioning, $\tau b = -.224$, $p = .006$. This indicates that the use of a binary framing has declined over the years, though it is still present.

In short, men and masculinities were largely taken for granted. When defined, "masculinity" was characterized in a range of ways: cultural norms and behaviours, identity and appearance, an attribute of cultures as a whole, a critique of traditional qualities associated with men and boys. While many have taken a binary framing, there appears to be a trend away from this and especially towards constructivist and pluralistic characterizations—even if the authors do not address this directly, such as with definitions or references to theoretical frameworks.

## 3.3  TYPES OF MASCULINITIES

I now review the types of masculinities invoked and defined in the literature. This is not an inclusive classification of masculinities, nor does each category comprise an exhaustive list of types. This represents the current state of affairs.

### 3.3.1  *General Types of Masculinities.*

Types of masculinities named across the corpus of papers, in order from most common to least:

- *Hypermasculinity*, described as "culture of college fraternities" ... "barbarian, manly hero" ... "dominance, violence, and lack of emotional expression" [21,25,58,82,99,126], with references to Witkowski [225] and Zolides et al. [232].
- *Hegemonic masculinity*, or the dominant, singular model of masculinity in a given society, which was not referenced, but used in several papers [2,99,118,190,227].
- *Toxic masculinity*, also not defined or sourced but used in several papers [25,112,142,146,168].
- *Normative masculinity*, referring to adherence to a given society's expectations for men and masculinities, described as "being assertive, demonstrating bravery through risk-taking, upholding heterosexuality and rejecting femininity, and establishing dominance through aggression" ... "appreciating and practicising sports" [168,211], with references to Mahalik et al. [127] and Pascoe and Bridges [152].
- *Alternative masculinities,* a pluralistic concept where "men are able to express their emotions, reject violence, and champion fighting all forms of oppression of women and other men" ... a "'softer' form of masculinity" [168,211], referencing Pascoe and Bridges [152].
- *Rugged masculinity,* described as "taming 'virgin' nature, the problems of habitation by indigenous peoples, and the issues of the supernatural associated with the encounter with wilderness" ... "'rugged individualism' culture in computing" [146,185], referencing Dourish [53], Ensmenger [57], Fox and Tang [65], and Salter and Blodgett [172].

Others include fragile masculinity [168], masculinity anxieties [168], male-default values [125] traditional masculinity [66], bapak (the Javanese version of hegemonic masculinity) [118], and supportive masculinity [173].

### 3.3.2  *Technology-Oriented Masculinities.*

The two most common forms of technology-oriented masculinities were:

- *Geek masculinity*, a general term variably described as a "masculine understanding of identity that is visible across technology culture" ... "in which technological mastery forms the basis of masculine esteem and social status" [25,67,136], with references to Kendall [105], Bucholtz [27], Eglash [55], and Lin and den Besten [119].
- *Toxic gamer culture*, focused on video game sites and which "frames gaming as a male-gendered, potentially violent space" [82,126], via Consalvo [39].

Others included brogrammers [25], alpha and beta masculinities [25], masculine technophile [190], technology czars and gurus [190], and masculine prototypicality in technology [162].

## 3.4   THEORIES AND FRAMEWORKS OF MASCULINITIES

Most (28, 92%) theories and frameworks referenced by authors in this corpus of work were general or non-disciplinary: identity intersectionality [71,91], construction [75,122], and multiplicity [28]; masculine norms [84,163], traits [17,179,186], roles [127], and cultures [166]; social construction [60,69]; performativity [37,213,214]; gender schema theory [18]; gender rules [210] and ideology [46]; Hofstede's masculinity index for cultures [90]; othering [144]; heteronormativity, heteropatriarchy [91], and value neutrality [9]; gender role strain paradigm [160,161]; father involvement [85]; muscle dysmorphia [157] and bigorexia [74]; male-dominated [65,93,226] and masculinized spaces [100]; autonomy from masculinity [170]; androcentrism and male-as-default [48]; alpha male effect [86]; and masculine disclosure [172]. Others (6, 18%) were tech-oriented: gender-agnostic platforms [125]; cultural stereotypes as gatekeepers [32] and men's/boy's clubs [99,169,201]; online disinhibition effect [207]; the manosphere [73]; co-production of gender and tech; and data feminism [49] (data as masculine, i.e., rational and objective).

## 3.5   SUMMARIZING APPROACHES TO MASCULINITIES AT CHI: SOCIAL, DESIGN, RESEARCH, CRITIQUE

I summarize the state of affairs across this 30-year corpus of CHI papers with the themes and codes in Table 1.

**Table 1:** *Thematic framework of approaches to masculinities at CHI*

| Theme | Code | Papers | Count (%) |
|---|---|---|---|
| Social | Behaviour | [2,3,5,6,11,14,15,20,25,36,45,47,51,52,58,62,66,68,76,80,83,94,97,103,109,114,117,123,125,126,128,133,134,136,138,140,142,146,147,149,153,154,158,162,164,167,168,170,177,180,181,183,190,192–195,199,200,202,208,211,217,218,221,223,227,228,230,231] | 69 (55%) |
| | Attitudes | [2–6,11,14,16,20,21,25,30,34,36,42–45,47,50–52,58,62,66–68,76,78,80,83,88,97,98,103,104,108–110,112,123–126,128,130–132,134,136,138,140,142,143,146–149,153,154,158,167,170,173,175,177,183,190,192,195–197,199,200,202,205,211,218,221,227,228,230] | 83 (66%) |
| | Identity | [2–6,11,16,21,25,31,33,36,51,52,58,66–68,70,76,79,83,94,98,99,103,110,117,125,126,128,134,138,140,142,146,147,153,154,158,177,188,190,192,195,197,199,200] | 50 (40%) |
| Design | Agent Attribute | [11,21,44,50,68,70,72,88,102,103,109,111,117,123,129,150,180,182,194,205,221,230] | 22 (17%) |
| | Interface Pattern | [3,14,30,61,66,72,78,104,108,110,131,132,134,137,138,143,147,150,167,171,202,211,217,229] | 24 (19%) |
| | Experience | [2–4,11,14,16,20,21,25,31,33,34,36,45,47,51,52,58,62,66,76,80,83,94,97,103,109,110,113,114,123,125,126,130,132–134,136,138,140,142,143,146,147,158,162,167,170,173,183,190,193,197,199,200,202,208,218,227,228,230,231] | 62 (49%) |
| | Space | [2,4–6,11,14,16,20,25,31,34,36,42,45,45,47,51,52,66,67,72,76,80,83,94,97,98,113,124–126,130,133,134,136–138,140,142,143,146,153,154,158,162,164,167,170,173,183,185,190,192,194,196,199,200,208,218,227,230,231] | 62 (49%) |
| Research | Method | [12,31,43,50,66,67,88,129,148,149,164,167,180,211] | 14 (11%) |
| | Reporting | [99,115,176,180,188,203,208] | 7 (6%) |
| Critique | Method | [4,23,30,42,72,88,89,99,128,148,154,167,175,177,180,181,188,195,223] | 19 (15%) |
| | Reporting | [23,89,99,109,126,154,180,188,223] | 9 (7%) |
| | Technology | [3,14,23,30,31,33,36,68,79,81,82,103,112,118,125,134,136,138,149,162,167,170,180,192,193,196] | 26 (21%) |
| | Field | [2,10,12,23,42,47,51,83,89,99,118,129,130,136,146,164,167,170,173,177,185,188,190,192,195,200,202,223] | 28 (22%) |

## 4  Discussion

Global shifts in how "masculinity" is viewed, personally and politically, are taking place alongside worldwide calls for action on gender bias and sexism in the technosphere. How has CHI risen to the challenge? While small, the pool of work focusing on or including a component of masculinities represents a diverse array of work. Still, there are gaps and potentials not yet traversed that may be especially suitably for HCI work, if not CHI specifically.

### 4.1  BRINGING IN MASCULINITIES FROM THE EXTANT LITERATURE

We can use the extant literature in men and masculinities studies and gender studies to seed new directions. Drawing on my expertise, I provide this curated list of influential work as a starting point.

- *Technomasculinity* [101] refers to how men portray themselves as advanced computer users and rely on this portrayal when relating to others. Here, computer use and especially mastery is taken on as a social identity. In HCI work, technomasculinity may guide: the selection of participants based on gender and/or technology-oriented identities, especially in multi-user contexts; the design of questionnaires and other probes that involve self-reports of technical identity, ability, and/or experience; and observational and analysis frameworks at sites and in data where technical mastery may play a role. For example, gendered self-selection of peer programming teams in a classroom setting may be understood through a technomasculinist framework. Technomasculinity relates to geek masculinity, found in this survey across several papers. Technomasculinity is about technical mastery as a form of power, while geek masculinity is an alternative when mastery of dominant forms of masculinity are perceived to be unachieved or unachievable.

- *Inclusive masculinity theory (IMT)* [7] was "developed to explain sport and fraternity settings where the social dynamics were not predicated on homophobia, stoicism or a rejection of the feminine" [8:549]. As this review has shown, technology spaces, activities, and roles across HCI contexts carry a range of masculine-centric or -dominant characteristics, values, and demarcations. Even the digital instantiation of sports, esports, has been explored [126]. If forms of masculinity trickle down from the larger culture, we may expect to find similarities when comparing to other masculine-centric or - dominant domains, like sports and fraternities. IMT presents a more nuanced framing to further shape these expectations and push us to consider alternatives. For example, we may be primed to look for certain forms of gendered interlocutions between a self-identifying jock and a stereotypically deferential feminine-voiced virtual assistant. We may miss or decentre engagements that do not fit expectations, such as the jock taking on emotional labour for a friend by searching about a seemingly non-gendered, benign topic, like the closest store to get a prepaid phone card.

- *Hybrid masculinities* are defined as "men's selective incorporation of performances and identity elements associated with marginalized and subordinated masculinities and femininities [24:246]. HCI now offers a wealth of ways in which to express, play with, challenge, deconstruct, and reify gender through identity and performance beyond traditional text modalities. We can choose and customize avatars; we can modify our appearance in realtime on Zoom; we can change our voice with vocalization software; we can even produce deepfakes of ourselves and others. A key element of this is *change*: we are not stuck with a single mode of expression. Hybrid masculinities could illuminate and explain longitudinal engagements with technologies as modes of self-expression. Social media, video games, wearables at cosplay events ... there are many HCI-oriented sites where hybrid masculinities could be found.

- *Caring masculinities* are defined by anti-domination, positive emotion, interdependence, and a focus on relations [56]. It may be employed as an alternative to the manosphere already explored at CHI [73]. Investigations could be as general as how people who identify as men or masculine conduct themselves in interpersonal exchanges online to specifically tracing out communities and movements centred on forms of caring masculinities, including and beyond anti-misogyny initiatives. We can also revisit the gaming and VR spaces found in the reviewed work to explore interactive narratives, characters, and mechanics that allow people to take on caring masculinist personas and modes of engagement.

- *Flexible, strategic* [13]*, and chameleon* [220] *masculinities* refer to "code-switching" in masculinity performance. Do those who identify as men or masculine switch between modes of expression, even suddenly or in rapid succession, when interacting with certain others or moving between technology spaces? Could such "code-switching" be embodied in virtual characters, agents, and robots designed with masculine cues? There is much to explore on either side of the HCI equation.

- *Postcolonial masculinities* are those that centre masculinities beyond Western and "First World" contexts [189]. In this review, only one was found: the Javanese bapak [118]. Initiatives in the field of HCI and especially at CHI have highlighted and pressed for recognition and engagement on matters of diversity, equity, and inclusion beyond gender. Could postcolonial masculinities complement the work this review has found on rugged masculinities [146,185], for example? This also means taking an intersectional lens to gender and in this case masculinities. For instance, do "caring masculinities" look the same in a South Asian WhatsApp group chat compared to a British one? We should be cautious about making assumptions and leaning on generalizations from WEIRD research at CHI [121] and adjacent spaces, such as HRI [234]. Still, HCI is a radical, creative, and political space, welcoming of inclusive knowledge and change in praxis. Ideas run the gamut. For instance, we could create Two-Spirit avatars and modes of engagement in interactive stories and video games that link masculinities, femininities, and gender identities and expression to sexuality and sociocultural roles and hierarchies that do not necessarily map on Western LGBTQI+ models and distinctions [235].

## 4.2 AN AGENDA FOR FUTURE INTERACTIONS WITH MASCULINITIES

I offer a non-exhaustive list of ideas and prompts based on the surveyed work and summarizing thematic framework.

### 4.2.1 *Social: Hack Masculinities with New Forms of Education and Events.*

Masculinities come to bear in social HCI contexts, with most work covered being attitudinal (66%), behavioural (55%), experiential (49%), and/or spatial (49%). A complementary, triangulated approach could be explored at these intersections in the forms of educational initiatives and social events. Plank [159] asks why we have tech events for girls but we do not have nursing events for boys. Could we design games that explore multiple forms of masculinities? What about VR applications that allow men and boys to freely explore gender expression? Could we design hackathons on technologies that confront toxic masculinity ... encourage caring masculinities ... or involve "feminine" activities, like digital sewing?

### 4.2.2 *Design: Create New Prototypes from the Lenses of Extant Theories.*

While the body of work covered in this review drew on several general (3.3.1) and HCI-oriented frameworks of masculinities (3.3.2), I point out several more candidates that have not yet been explored (4.1). How can we use these theories in HCI? The paths forward are too numerous to name, but I can offer a few more ideas. How could HCI approach, for example, postcolonial masculinities, especially given the decolonizing work [154,206] already underway? Are there caring masculinities in the e-health space or in mental health and wellbeing spaces on social media? Could there be?

### 4.2.3 *Research: Explore a Diversity of Theories.*

As the content analysis results (3.2) indicate, much of the work covered in this review appears to speak to theories, even if those theories were not used. Can we revisit previous work—even just the data for further analysis—in case instances of novel (for CHI) forms of masculinities were missed? Can we offer our data as part of a new open science initiative to uncover implicit connections to existing theoretical frameworks? Doing so would not only be useful for the field of HCI, but also feed back into larger knowledge and theoretical bases of gender and masculinities. This could show how CHI as a venue contributes to general knowledge in a concrete way. This would also reveal the how HCI is distinct, leading to offshoot theories and potentially new ideas for design and research.

### 4.2.4 *Critique: Let's Be Reflexive and Change Our Ways.*

We lean on masculinities, whether we are conscious of it or not. Still, only 3% of the corpus operationalized "masculinities," with 64% leaving the concept undefined and nearly half (48%) taking an unclear stance, even for recruiting and/or demographics reporting. Still, the other half (48%) has taken on a pluralistic approach, even without providing a clear definition or using a theoretical framework. We may be at a key juncture for reflexivity [164] as a field of study and practice. CHI can lead the way. Change can be small, such as rethinking how we ask about gender for demographics [187]. How do we wish to interact with masculinities?

## 4.3  LIMITATIONS

This work was limited by the focus on the CHI conference; future work should scope out the literature in other HCI venues, including conferences, journals, and other venues. I alone carried out this work; while aimed for self-correction by remaining reflexive and employing the PRISMA-ScR, I acknowledge that the codings and classifications could be limited by this solo approach. Future work can test and expand these frameworks with multiple raters.

## 5   Conclusion

In this scoping review, I have traced out a history of masculinities at the premier international CHI conference. I have extended the base offered by this body of work by weaving in extant theories and highlighting trajectories for future work. HCI spaces like CHI have much to offer, in the past, present, and future, by approaching masculinities as a matter of inclusion, diversity, and social justice for everyone.

### ACKNOWLEDGMENTS

This work was funded by a Tokyo Tech Young Investigator Engineering Award.